\documentclass[11pt,twoside]{article}
\usepackage{asp2010}
\usepackage{natbib}
\usepackage{graphicx}
\usepackage{epstopdf}
\usepackage{auto-pst-pdf}
\resetcounters

\begin{document}

\title{Interferometric imaging diagnostics of  X Hya's circumstellar environment}
\author{X. Haubois$^1$, M. Wittkowski$^2$, G. Perrin$^1$, P. Kervella$^1$, S.T.  Ridgway$^3$ and E. Thi\'ebaut$^4$}

\affil{$^1$ LESIA, Observatoire de Paris, CNRS UMR 8109, UPMC, Universit\'e Paris Diderot, 5 place Jules Janssen, F-92195 Meudon, France} 
\affil{$^2$  European Southern Observatory (ESO), Karl-Schwarzschild-Str. 2, D-85748 Garching bei M\"{u}nchen, Germany} 
\affil{$^3$ Kitt Peak National Observatory, National Optical Astronomy Observatories,  P.O. Box 26732,  Tucson, AZ 85726-6732 U.S.A.} 
\affil{$^4$  Centre de Recherche Astrophysique de Lyon, CNRS/UMR 5574, 69561 Saint Genis Laval, France} 

\begin{abstract}

Optical interferometry is a powerful tool to investigate the close environment of AGB stars. With a spatial resolution of a few milli-arcseconds, it is even possible to image directly the surface of angularly large objects. This is of special interest for Mira stars and red supergiants for which the dust-wind is initiated from or very close to the photosphere by an interplay between pulsation and convection. Based on two-epoch interferometric observations of the Mira star X Hya, we present how the variation of the angular size with wavelength challenges pulsation models and how reconstructed images can reveal the evolution of the object shape and of its asymmetric structures. 

\end{abstract}

\section{Introduction}

The mass loss that determines the fate of AGB stars is triggered by convection and pulsation phenomena. The subsequent formation and transport of material from the surface to the interstellar medium still remain to be discovered.
The very high resolution that interferometric imaging delivers allows to observe the surface, the molecular layers as well as the dusty atmosphere. In particular, asymmetries in the brightness distribution, e.g. spots on the surface or clumps in the atmosphere, are the signatures of processes such as convection or pulsation. The key to pin down their nature and origin is to derive their spectral dependence and their evolution timescales. These properties are guaranteed to bring decisive insights on the  mass loss processes at play in these stars. With this aim, we here report on interferometric observations of the mira star X Hya observed with AMBER.

\section{Spectro-interferometric observations}
X Hya is a M7e Mira star with a 301 day pulsation period located at a distance of $\sim 440$ pc \citep{2008MNRAS.386..313W}.  We observed it with AMBER \citep{2007A&A...464....1P} at two epochs corresponding to two phases of its pulsation cycle: $\Phi =0.0$ and $\Phi =0.2$.  A multi-wavelength observation is required to distinguish the counterpart of the molecular layer compared to the stellar photosphere. We thus performed J-H-K observations (covering from 1.17 $\mu$m to 2.37 $\mu$m) at low spectral resolution (R $\sim 30$). 

\begin{figure*}[!h]
 \centering
\includegraphics[width=2.6in]{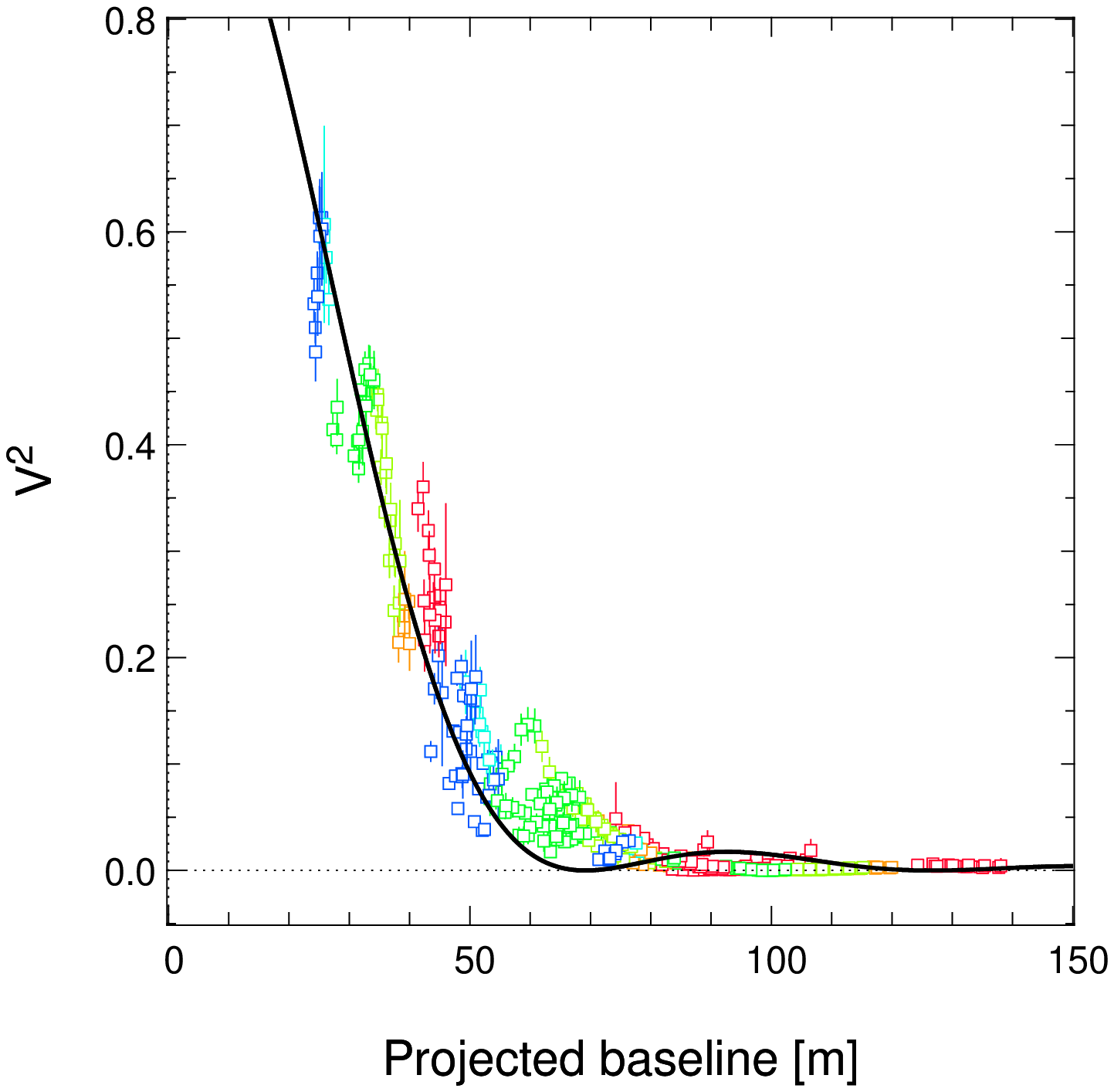}
\includegraphics[width=2.6in]{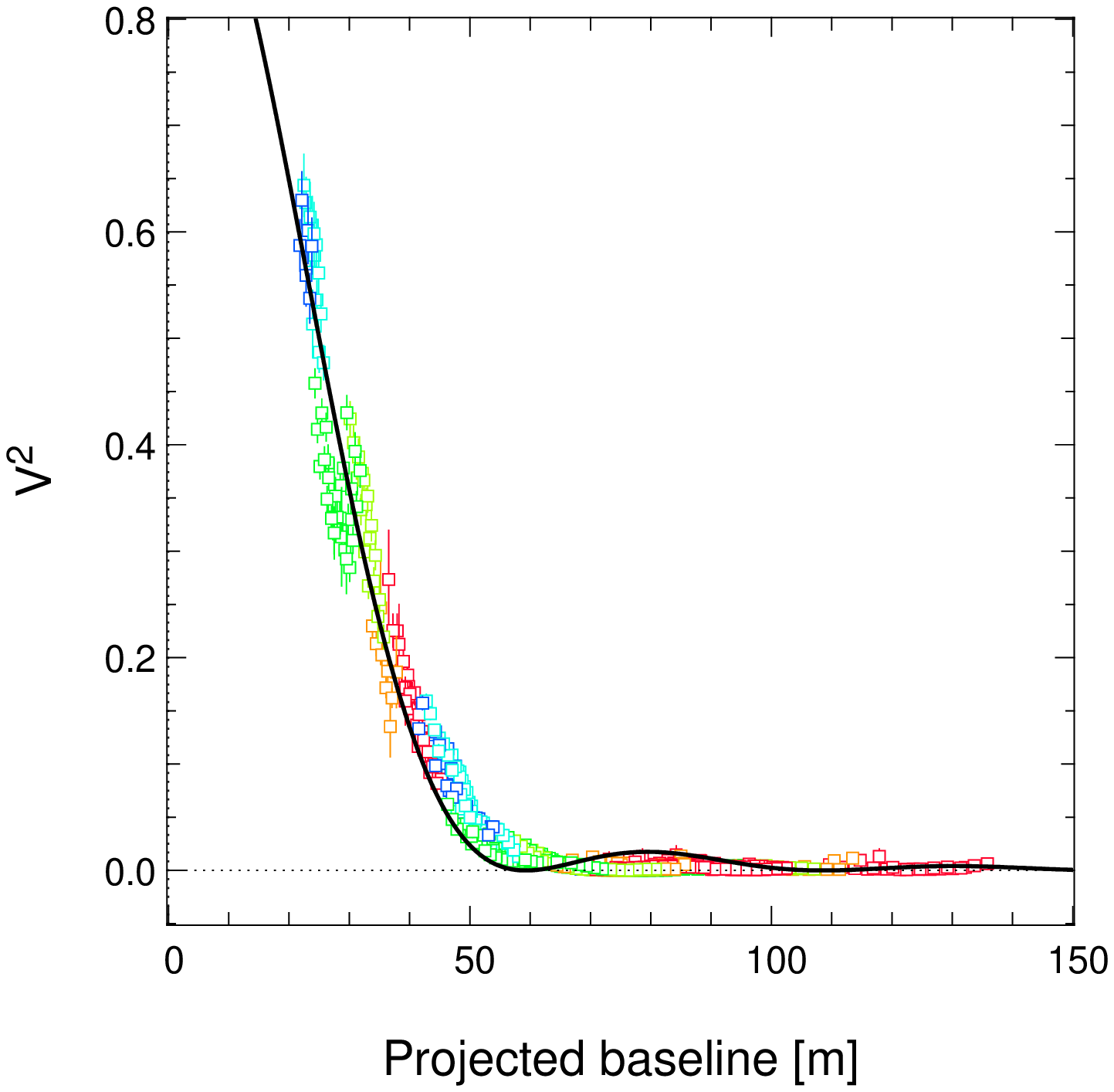}
\caption{Squared visibilities data at Epoch A (left) and Epoch B (right). Colors indicate the 6 different spectral bands. The solid lines represent a 6 mas (for Epoch A) and a 7 mas (for Epoch B) uniform disks at 1.75 microns (average wavelength of the covered spectral interval).
}
       \label{v2data}
 \end{figure*}

The datasets we obtained at each epoch are rich and complex. Even at short baselines, squared visibilities ($V^{2}$) show a significant departure from a simple centro-symmetric geometry such as an uniform disk model (Fig.~\ref{v2data}). In order to understand the spectral and morphological signatures on the $V^{2}$ data, we first divided the data in six spectral bands as shown on Tab.~\ref{tab:bands}. This division is based on an analysis of the visibility variation with the wavelength and is coherent with a previous spectro-interferometric analysis of X Hya \citep{2011A&A...532L...7W}.

We first performed a uniform disk (UD) model to get a first estimate of the angular size of the object in the different bands and at the different position angles (PAs) we observed.  We only considered squared visibility data whose spatial frequencies belonged to the first lobe in order to avoid any contamination from limb darkening or structures smaller than the stellar diameter. An average of the measurements over the PAs is shown in Table~\ref{tab:bands}.

\begin{table}[h!]
\centering
\begin{tabular}{cccc}
\hline
 Wavelength range & Contribution &  $\phi_{UD}$  & $ \phi_{UD}$ \\  
  & & $\Phi =0.0$ & $\Phi =0.2$ \\ \hline
$1.1-1.4 \,\rm\mu m  $  & Continuum (1) & 5.94 $\pm$ 0.15\,\rm mas  & 7.41 $\pm$ 0.40\,\rm mas \\ \hline
$1.4-1.5 \,\rm\mu m  $ &  $\rm H_{2}0$ (1)  & 6.90 $\pm$ 0.13\,\rm mas & 7.85  $\pm$ 0.49\,\rm mas \\  \hline
$1.5-1.7 \,\rm\mu m  $ & CO (1)  & 6.03 $\pm$ 0.59 \,\rm mas& 7.19 $\pm$ 0.85 \,\rm mas \\  \hline
$1.7-2.1 \,\rm\mu m$ &  $\rm H_{2}0$ (2) & 6.52 $\pm$ 0.23 \,\rm mas&   8.24 $\pm$ 0.86 \,\rm mas \\  \hline
$2.1-2.25 \,\rm\mu m $ & Continuum (2) & 6.20 $\pm$ 0.35 \,\rm mas & 7.02 $\pm$ 0.86 \,\rm mas \\  \hline
$2.25-2.4 \,\rm\mu m $ & CO (2)  & 6.80 $\pm$ 0.42\,\rm mas&  7.54 $\pm$  0.79 \,\rm mas\\ \hline
\end{tabular}
\caption{\label{tab:bands} Division of the covered spectral range in 6 spectral bands and uniform disk diameters for the two observing epochs.}
\end{table}

\subsection{Radiative transfer modeling}
To better understand the spectral dependence of this complex dataset, we used a grid of dynamic model atmosphere series computed with the P/M and CODEX codes. These models develop a self-excited pulsation mechanism and sample opacities over 4300 wavelengths running from 200 nm to 50 $\mu$m, with 1 nm sampling between 200 nm and 3 $\mu$m. For further details see \cite{2008MNRAS.391.1994I,2011MNRAS.418..114I}.
 
\begin{figure*}
 \centering
 \includegraphics[width=4.5in]{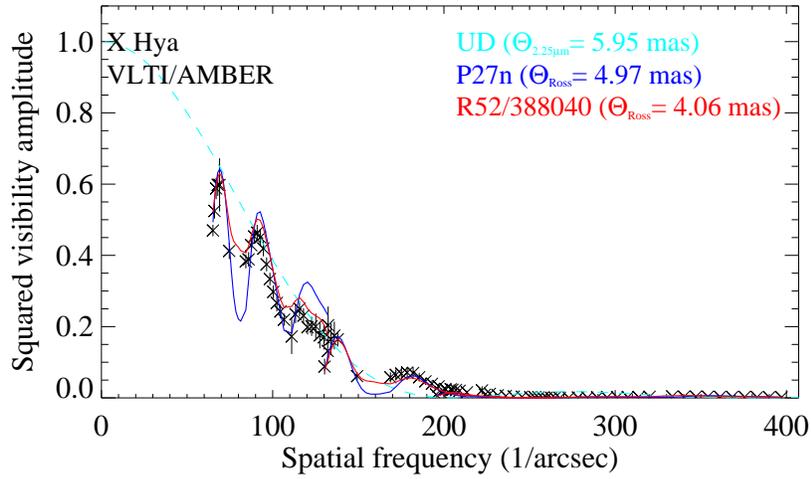} 
\caption{Modeling of the squared visibility data (in black) with CODEX/R52 (red), P (in blue) and UD (cyan) models. The Rosseland diameter indicated in the legend was derived from the best-fit of these models.}
       \label{graph_CODEX} 
 \end{figure*}

For a set of parameters, these models produce radial profiles of the intensity across the chosen spectral band that are turned into $V^{2}$ by a Fourier transform. These latter were finally fitted to our data by adjusting the Rosseland diameter and the pulsation phase (for CODEX models only). Figure~\ref{graph_CODEX} shows $V^{2}$-fitted curves for one baseline. At the difference of the simple geometric models, these physical models that take into account $\rm H_{2}0$ and CO opacities, do reproduce the variation of the $V^{2}$ with the spatial frequency fairly well. The results of the model parameters fitting are summarised in Tab.~\ref{tab:CODEX}.

\begin{table}[h!]
\centering
\begin{tabular}{ccc}
\hline
Models & Epoch A, $\Phi =0.0$  & Epoch B, $\Phi =0.2$\\  
\hline
Best P model & P27n  & P14n  \\
$\theta_{Ross}$ (mas) & 4.95 & 7.38  \\
$\chi_{red}^{2}$ & 11.5 & 62.1 \\
\hline
Best M model & M16n  &  M12\\
$\theta_{Ross}$ (mas) & 5.00 & 6.58    \\
$\chi_{red}^{2}$ & 14.3 & 63.2 \\
\hline
Best R52 model & 388040 & 386280 \\
$\theta_{Ross}$ (mas)  & 4.02 & 4.54 \\
$\chi_{red}^{2}$ & 8.1 & 44.3\\
\hline
Best o54 model & o54/262160 & o54/287940  \\
$\theta_{Ross} (mas)$  & 4.76  & 5.86  \\
$\chi_{red}^{2}$ &  7.4 & 41.2 \\
$\Phi_{CODEX}$& 0.7  & 0.5 \\
\hline
\end{tabular}
\caption{\label{tab:CODEX} Results of the fitting of radiative transfer models. }
\end{table}

\section{Image reconstruction}
In the previous sections, we only took into account the $V^{2}$ data which means the object was thought to be centro-symmetric. However, because AMBER recombines the beams from three telescopes, we do have a phase information called closure phase that is a measurement of the asymmetry of the object.  The closure phase signals exhibited a significant departure from the null value in all spectral bands.

In order to reconstruct the spatial intensity distribution out of such a complex dataset of $V^{2}$ and closure phases, it is mandatory to use image reconstruction algorithms such as MIRA \citep{2008SPIE.7013E..1IT}. However, given our limited UV coverage in individual bands, we can't aim at reconstructing a model-independent image. Nevertheless, we carried out an image reconstruction approach based on our parametric modeling results. We therefore used 1D intensity profiles from the radiative transfer models that we turned into a prior image. The reconstructed image was obtained by adjusting the image to the data and by regularizing towards the prior image.

\begin{figure*}[!h]
  \includegraphics[width=2.6in]{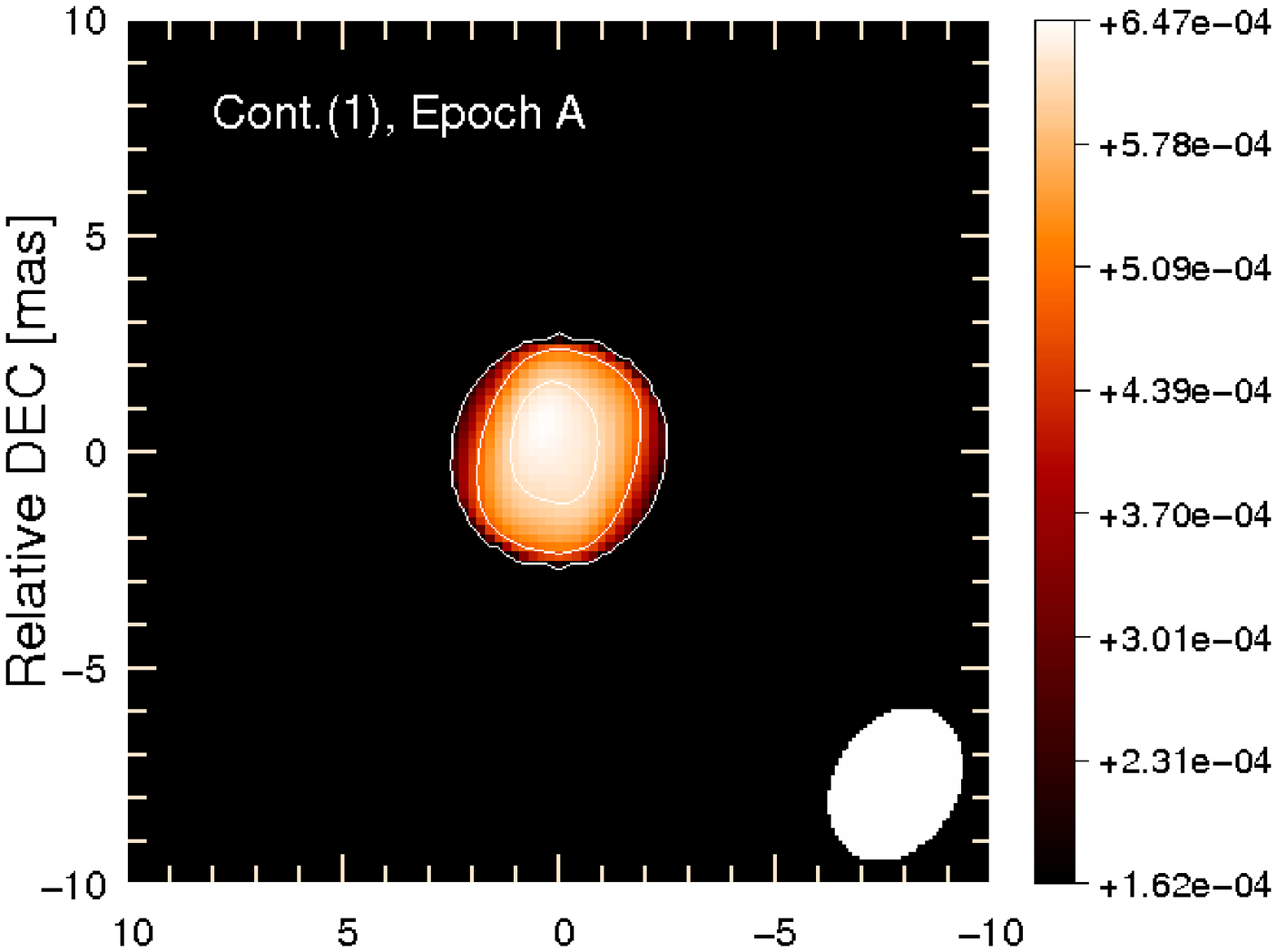}
  \includegraphics[width=2.6in]{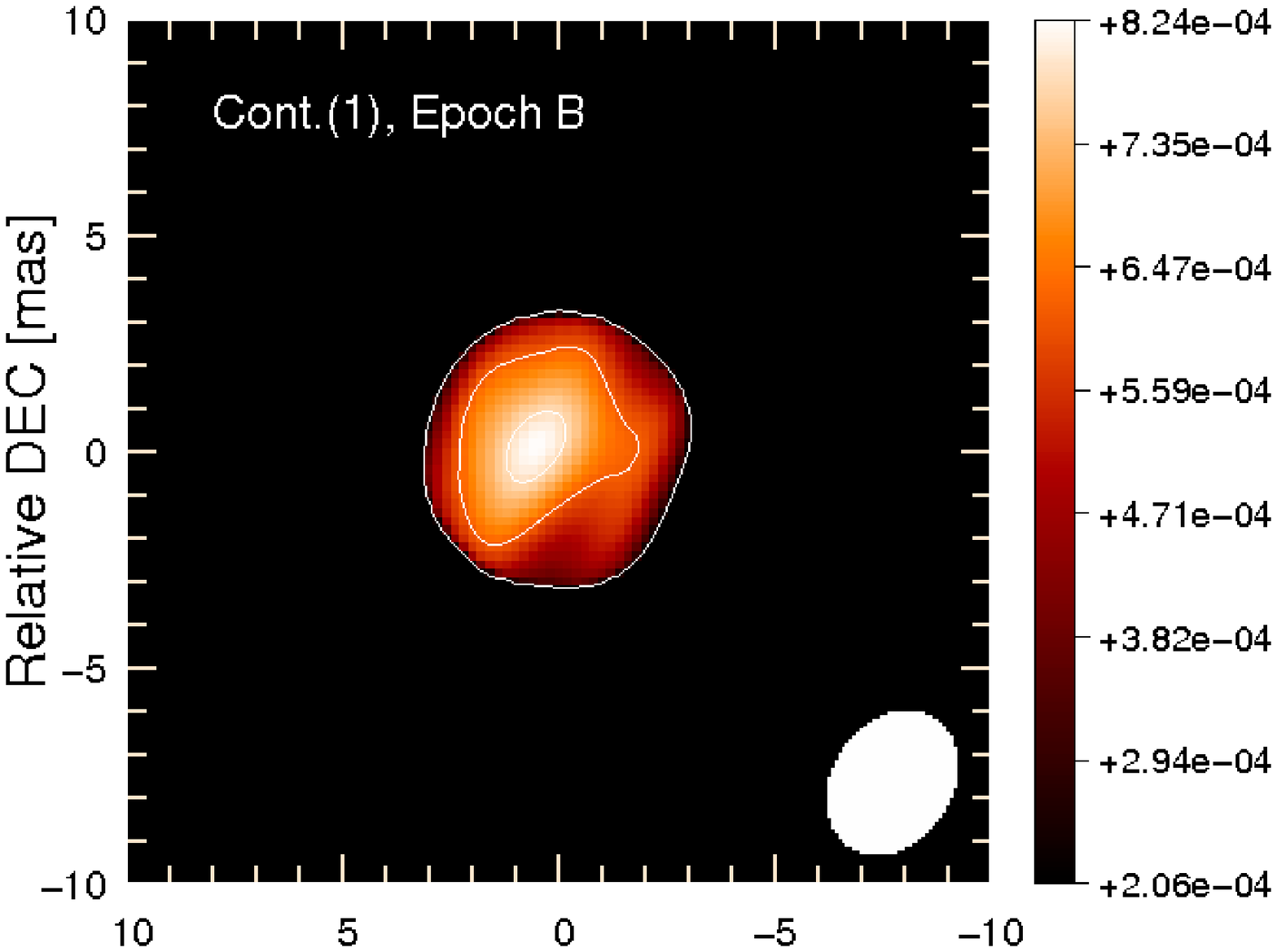}
    \includegraphics[width=2.6in]{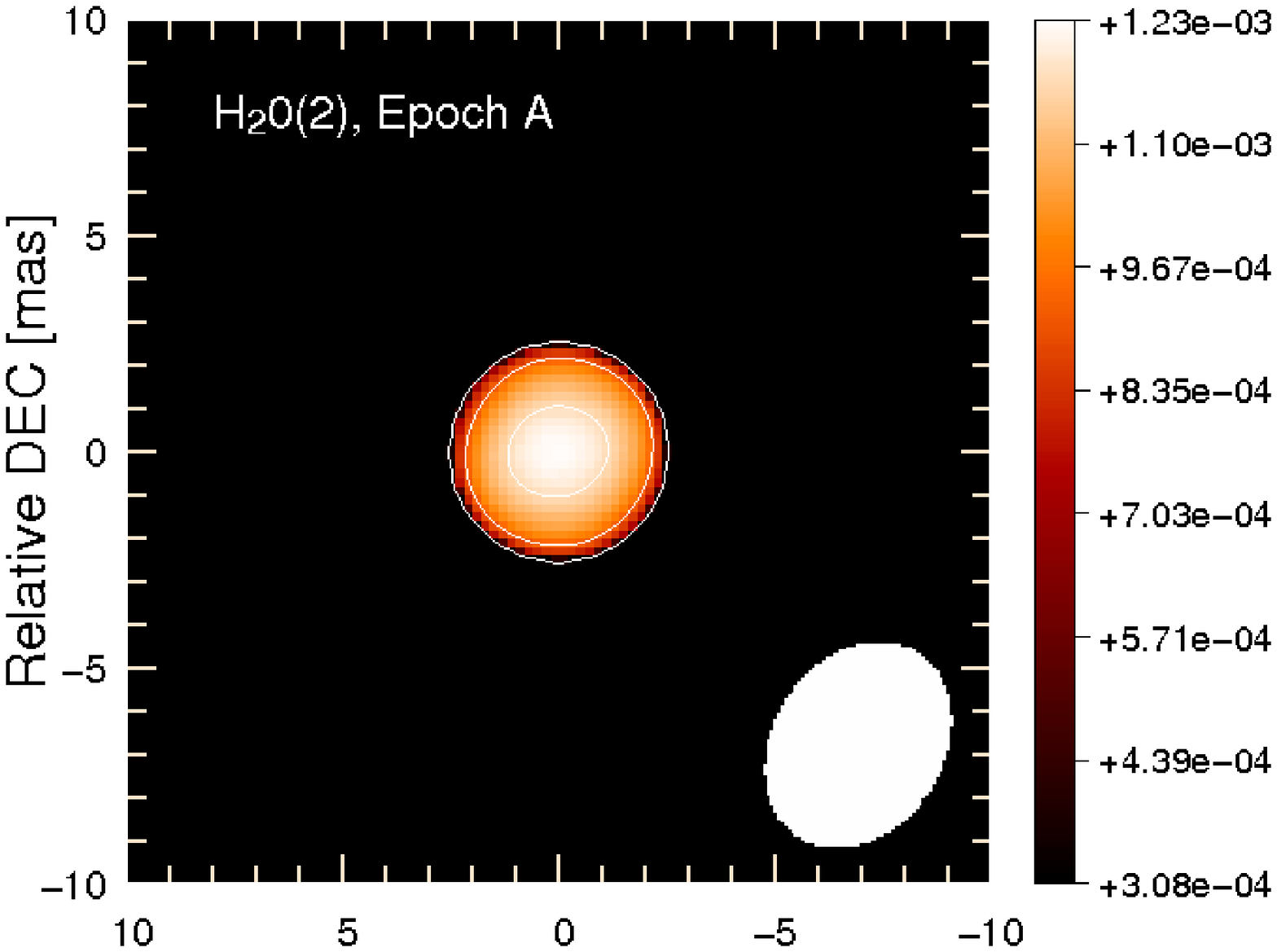}
   \includegraphics[width=2.6in]{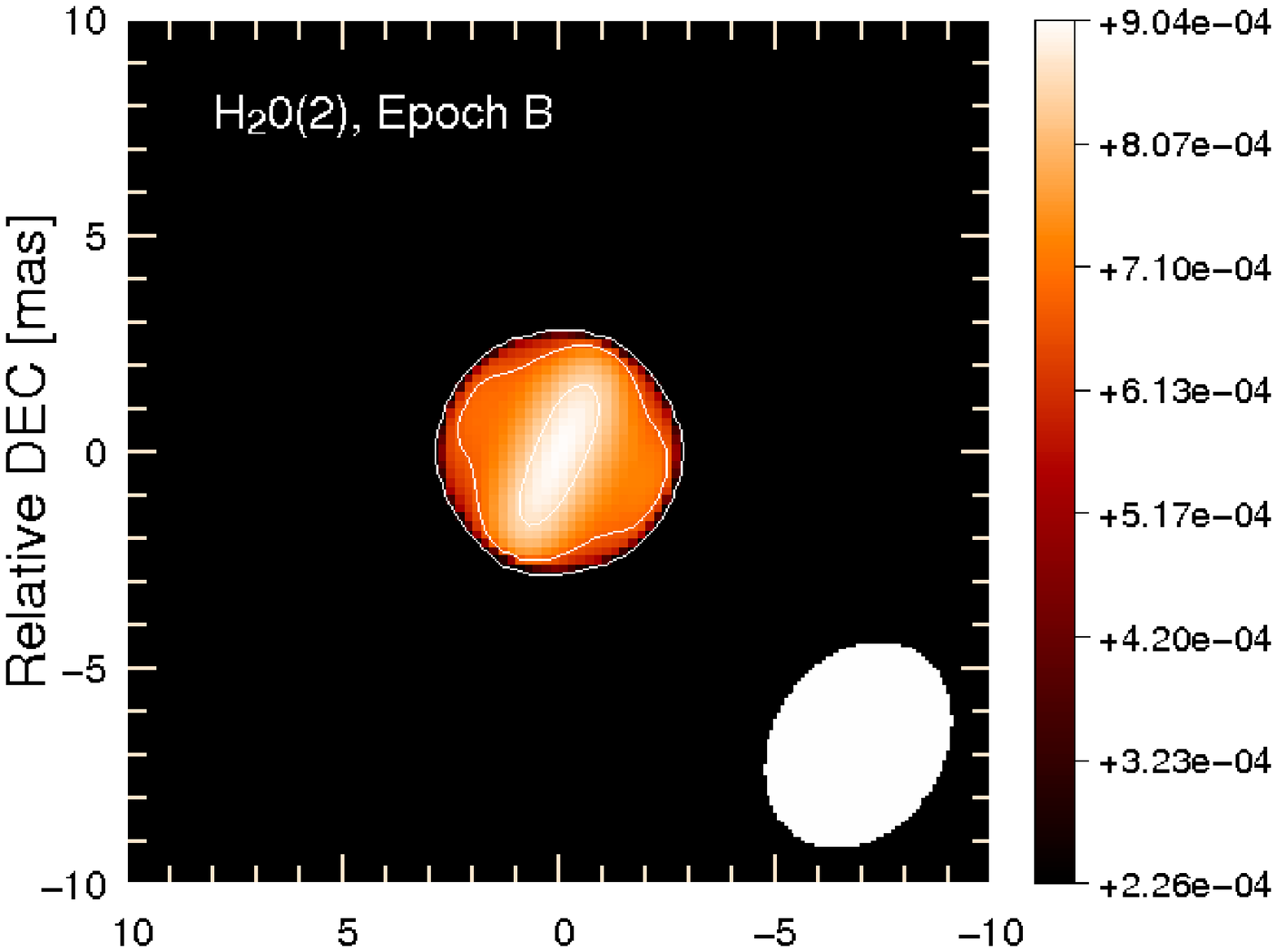}
  \caption{Reconstructed images with MIRA for spectral bands "Continuum 1" and "$\rm H_{2}0$ (2)" at Epoch A (left column) and Epoch B (right column).  The interferometric beam size is displayed in the bottom-right corner of each image. White contours mark areas of 25\%, 75\% and 95\% of the maximum flux of the image.}
     \label{images1}
 \end{figure*}

Several regularization techniques were used. The best results were obtained with a quadratic regularization towards an a priori object image derived from the R52/388040 model intensity profile. All the reduced $\chi^{2}$ values associated with the best reconstructed images are lower than 1. Figure~\ref{images1} shows images for two spectral bands at the two phases.
 
  In agreement with our previous results, the overall size of the object increased between the two epochs in all spectral bands. X Hya also presents a more complex morphology at the second phase and especially in the continuum and in the $\rm H_{2}0$ bands. In order to push further the interpretation of these observations, we plan to compare these images with full 3D hydro-radiative transfer models that also include self-excited pulsating atmospheres.

\section{Conclusion}
We report on low-resolution spectro-interferometric AMBER observations of the mira star X Hya. After a first simple geometric analysis, we performed a model-fitting with the P/M and CODEX atmosphere series that were able to reproduce the chromatic dependence of the squared visibilities. In order to adjust the non-zero closure phase signals that we observed in all spectral bands, we used the reconstruction algorithm MIRA to obtain images at the two phases of the pulsation cycle. We observed a higher degree of asymmetry at the second epoch ($\Phi =0.2$) in the continuum and in the $\rm H_{2}0$ bands.  We will compare these reconstructed images with 3D hydro-radiative transfer models of pulsating atmospheres.
\bibliography{aspauthor}{}
\bibliographystyle{asp2010}

\end{document}